**Correlation between electric field-induced phase transition and piezoelectricity in lead zirconate titanate films**

V. Kovacova[1], N. Vaxelaire[1], G. Le Rhun[1], P. Gergaud[1], T. Schmitz-Kempen[2], E. Defay[1]

1. CEA, LETI, MINATEC Campus, 17 rue des Martyrs, 38054 GRENOBLE Cedex 9, France.

2. AixACCT Syst GmbH, D-52068 Aachen, Germany

**Abstract**

We observed that electric field induces phase transition from tetragonal to rhombohedral in polycrystalline morphotropic lead zirconate titanate (PZT) films, as reported in 2011 for bulk PZT. Moreover, we evidenced that this field-induced phase transition is strongly correlated with PZT film piezoelectric properties, that is to say the larger the phase transition, the larger the longitudinal piezoelectric coefficient $d_{33,eff}$. Although $d_{33,eff}$ is already comprised between as 150 to 170 pm/V, our observation suggests that one could obtain larger $d_{33,eff}$ values, namely 250 pm/V, by optimizing the field-induced phase transition thanks to composition fine tuning.

Although efficient lead-free piezoelectric thin films have been recently proposed in the literature [1-4], lead-based films as Pb(Zr,Ti)O$_3$ (PZT) [5,6] or Pb(Mg,Nb)O$_3$-PbTiO$_3$ (PMN-PT) [7] always exhibit the best piezoelectric properties. More specifically, they are strongly enhanced at the so-called morphotropic phase boundary (MPB) [8]. For PZT, the MPB is at the interface between the tetragonal and rhombohedral phases while Zr/Ti atomic ratio is close to 52/48. In addition to these phases, the monoclinic phase might be present in the MPB, as evidenced in 1999 by Noheda et al. [9]. This phase coexistence and consequently the polarization mobility increase are thought to be the reason why piezoelectricity is enhanced in the MPB region. In 2011, Hinterstein et al. showed by *in situ* observations that bulk ceramic MPB PZT experiences structural changes while an electric field is applied [10]. More specifically, they reported that the application of an electric field reveals an increase of the monoclinic phase fraction. As they observed that most of PZT was composed of tetragonal and monoclinic phases, it strongly suggests that PZT experienced a tetragonal-to-monoclinic phase transition under electric field, and not only rotation of the polarization. An open question is therefore whether there is a correlation between large piezoelectric effects and this field-induced phase transition. Today, it is widely believed that most of the piezoelectric effect is induced by domain walls switching. It is the so-called extrinsic contribution [11]. Several groups showed that PZT bulk and thin films in the tetragonal phase experience *a* to *c* domain switching while electric field is applied [12-15]. In this paper, we show *in situ* structural modifications in MPB PZT films versus electric field and compare with their piezoelectric properties. We aim to correlate field-induced phase transition and piezoelectric properties. This correlation could have a strong impact on piezoelectric material design for applications, as inkjet devices, integrated optical lenses or micro-actuators in general [16].

In this study, we prepared sol gel MPB PZT thin films. We characterized their piezoelectric coefficient $d_{33,eff}$ and X-Ray diffraction (XRD) measurements were performed revealing PZT microstructure modifications while dc-electric field was applied.

The substrates are 200mm in diameter and made of Pt(111) 100 nm/TiO$_2$ 20 nm/SiO$_2$ 500 nm/Si 750μm. Thermal SiO$_2$ is grown at 1100°C in oxygen. TiO$_2$ is obtained by depositing 10nm of Ti followed by thermal annealing at 700°C in oxygen for 30 min. Pt bottom electrode is sputtered at 450°C. Each layer of PZT is spun, dried at 130°C and calcinated at 360°C. The wafer is annealed at 700°C for 1 min each time six layers have been deposited. This step enables to crystallize PZT in the desired perovskite structure. Each single crystallized PZT layer is 60 nm-thick. Eventually, PZT thin film is 1 μm-thick. As proposed by Calame et al. [5], the built-in composition gradient appearing in PZT sol gel can be dramatically decreased by using successive precursors with different compositions in order to decrease the composition variations. We have implemented this method with three different Zr/Ti ratio, namely 59/41, 52/48 and 43/57. Finally, the top electrode is made of 100nm-thick sputtered Ru. The designs realized in this study only involve Ru patterning to obtain the desired devices. Once resist has been deposited, exposed and developed, Ru is wet etched with NaOCl. Finally, the resist is removed with acetone.

The piezoelectric coefficient $d_{33,eff}$ has been measured with Aixacct Double-Beam Laser Interferometry (DBLI) apparatus [17]. It requires polishing the substrate back side to insure a reflecting surface. The DBLI resolution is better than 1 pm/V. Therefore, the experimental accuracy is well below 1% as the measured deflections in this paper are in the nm range.

In order to achieve *in situ* XRD measurements, 7 x 7 mm² samples with one 5 x 5 mm² squared top electrode were cut out from the prepared wafers. Samples were then glued on a small piece of Printed Circuit Board (PCB). Silver paste was spread on one cleaved side of the sample to contact the bottom electrode. 38 μm-diameter gold wires were glued with silver paste between the sample electrodes and the PCB. Stronger isolated wires were then soldered between the PCB and the voltage supply.

θ/2θ symmetrical measurements were performed on a D5000 Bruker diffractometer in a Bragg-Brentano configuration (BB). The *in-operando* XRD measurements were achieved with a four circle X-Pert Panalytical diffractometer using a poly-capillary lens as primary optics and a crossed slit collimator with an aperture of 2 x 2 mm². Taking into account the beam divergence, the beam size on sample is therefore about 3 x 4 mm². The secondary optic is a parallel plate collimator with a flat-graphite monochromator and a proportional detector. In both systems the X-ray source is a copper anode and a 1°-offset has been applied in order to avoid the strong Si substrate reflections.

In order to define the samples' texture and phases present in PZT, pole figure measurements on planes {100}, {110}, {111} together with θ/2θ scans from 20° to 102° were performed. The voltage impact on PZT was subsequently studied. For each voltage applied, longer acquisition θ/2θ symmetrical scans were run around $\{100\}_t/\{100\}_r/\{001\}_t$ and their upper reflections orders, namely $\{200\}_t/\{200\}_r/\{002\}_t$ and $\{400\}_t/\{400\}_r/\{004\}_t$. $_t$ and $_r$ stand respectively for tetragonal and rhombohedral. The 2θ step width and duration were respectively 0.02° and 2s for {100} and {200} planes. The time per step for {400} planes was 4s. The 2θ range was [20.3, 23.3]°, [43, 46]° and [95, 101]° for the first, second and fourth order of (100) reflection respectively. The successive voltages applied to the sample were 20V, 10V, 0V, -10V, -20V, -10V, 0V, 10V, 20V and 30V. Note that no pre-poling has been performed prior to this experiment.

The PZT XRD profiles were analyzed using two crystallographic space groups, namely tetragonal P4mm and rhombohedral R3m as well as Ruthenium hexagonal P63/mmc. MAUD software [18] has been used to fit the experimental results. Assuming (100) fiber texture for both P4mm and R3m structures, the model has enabled to quantify the phase percentage, the texture of the tetragonal phase and eventually the strain in the rhombohedral phase. The crystallographic models used in the simulation were "Polynomial" for background modeling, "Standard function" for P4mm and R3m texture and "Isotropic size-strain" for crystallite size and micro-strain.

Measured $d_{33,eff}$ at several positions on the sample are all in between 120 and 130 pm/V. This value is in line with the best values obtained for gradient free PZT as reported in Calame's thesis [19].

The BB diffraction experiments are shown in Fig. 2. Perovskite is the only visible phase, without any trace of pyrochlore. Fig. 2 a., b. and c. shows the pole figures for planes {100}, {110} and {111}. As the microstructure varies in ψ angle and is independent of φ angle, it proves that PZT has a strong $(100)_{r/t}$ fibre texture with a mosaicity of 6°. Although not represented here, rocking curves under different voltages on {200} planes show that mosaicity remains constant with electric field. Because these films exhibit a strong $(100)_{r/t}$ fibre texture, the constant mosaicity induces that the subsequent BB analysis of planes {n00} performed along the fibre axis at different electric fields enable to observe the behaviour of nearly all the crystallites at once.

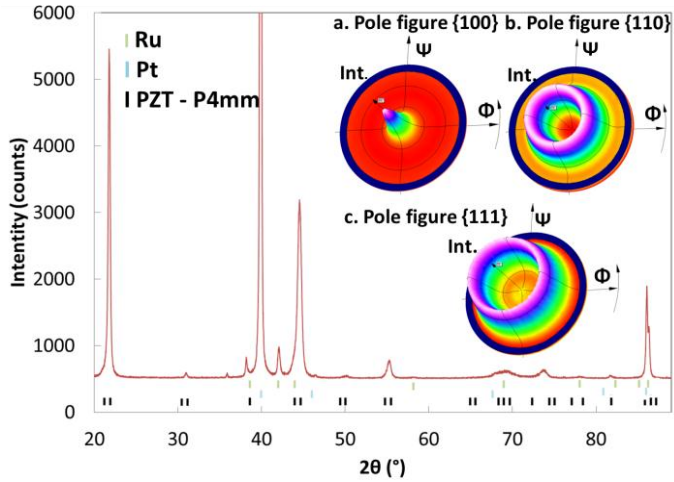

*Fig. 2: ϑ/2ϑ profile and pole figures of PZT a. {100} pole figure, b. {110} pole figure and c. {111} pole figure.*

Fig. 3 shows the BB diffraction profiles around (400) peaks at different voltages. The position of the (400) peaks for P4mm and R3m phases are also reported after the Inorganic Crystal Structure Database (ICSD) - 92059 for P4mm, 24562 for R3m-. Only {400} reflections are exploited since {100} and {200} reflections are at too low Bragg angles to be distinctly resolved. Although one should expect three peaks in this range, we only observe one experimental peak with contributions coming from P4mm and R3m phases as detailed later. This peak shifts towards lower angles when voltage is applied. The peak amplitude decreases with voltage. The amount of (004) P4mm phase, that is to say tetragonal c-domains, is very weak and does not change significantly with voltage. Although it is not visible in Fig.3, the peaks position versus the applied voltage is reproducible, as will be shown later in Fig.5.

*Fig. 3: ϑ -2ϑ profile evolution of PZT around peak (400) vs different voltages. The positions of (400) peaks of P4mm and R3m phases are also indicated. The arrow underlines the peak behaviour with voltage, namely low angle shift plus amplitude decrease.*

Fig. 4 shows the peak profile refinement of the θ-2θ scan achieved with MAUD in the vicinity of peaks $(400)_t$ and $(400)_r$ at 0 and 30V. As already stated, only P4mm and R3m phases were identified in PZT. This is true for all applied voltages as the peaks area is constant versus voltage, which means that no other phase appears with voltage. Ru (202) peak is also shown in this area of interest. The $(004)_t$ signal is not visible whatever the voltage. Consequently, c-domains have not been considered in the refinement process. Ru peak together with background have been fitted at 0V and then kept constant for all other refinements. These limitations have not hindered a very satisfactory fit of all experimental data, as depicted in Fig.4. At 0V, P4mm and R3m phases are clearly identified. At 30V, the amount of R3m drastically increases and cancels out P4mm occurrence. Note that there is no need to implement strain in the model to fit the experimental data.

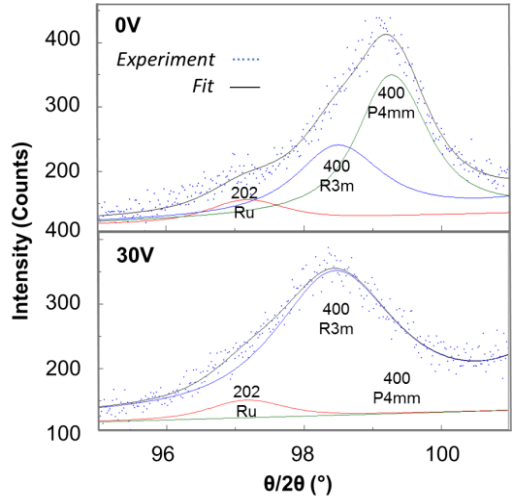

*Fig. 4: Refinement at 0V and 30V of ϑ-2ϑ scans of the PZT film. (400)$_r$, (400)$_t$ and Ru (202) reflection have been identified and fitted. The dots are experimental data, the black line corresponds to the sum of 202 Ru (red), 400 R3m (blue) and 400 P4mm (green) fits.*

The relative volume proportion of phases P4mm and R3m versus voltage has been determined with MAUD and is shown in Fig.5. Hence R3m stands for 40% of the phase volume at 0V and 100% at 30V. This effect is approximately symmetrical when applying negative voltage. The experimental reproducibility has been used to assess error bars. It notably shows that there is no or very little hysteresis versus voltage.

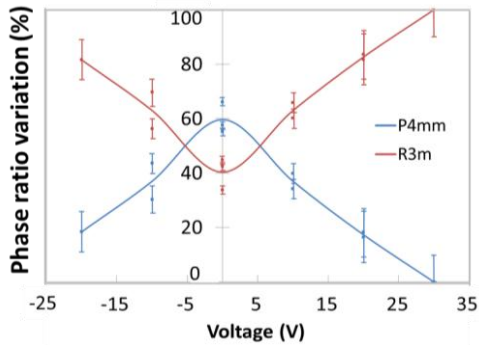

*Fig. 5: phase variation versus applied voltage of PZT P4mm and R3m phases in the film. Solid points are the results of the profile refinement. There are two points for all voltages except at -20V and +30V. Solid lines are guides for the eyes, based on the solid points' average value. Error bars are based on the reproducibility obtained at the same voltage.*

As P4mm and R3m are the only phases, the increase of R3m amount with voltage magnitude means that P4mm volume decreases accordingly, that is to say there is a voltage-induced phase transition. More specifically, as we observed that the appearance of tetragonal c-domains is not visible, there is a phase transition from a-oriented P4mm to R3m. The transition of 52/48 PZT between two phases induced by electric field has already been reported by Hinterstein et al. [10] in bulk ceramic. They actually identified a transition from P4mm towards monoclinic Cm. In our study, R3m fits better than Cm our results. Note that Cm is very similar to R3m, and has been stated as rhombohedral-like phase by Hinterstein.

Besides, the fit gives the opportunity to roughly extract $d_{33,eff}$ independently from the DBLI method. Indeed, we know 1) the amount of volume phase that has transited from P4mm to R3m (60% at 30V) and 2) the peaks position (98.2° for R3m and 99.2° for P4mm). This method gives $d_{33,eff}$ = 120 +/- 25 pm/V at 30V, which is in the range of what has been measured with the DBLI method (120-130 pm/V). Therefore, it proves that the piezoelectric effect of this MPB gradient free PZT film is mainly due to the voltage-induced phase transition from tetragonal to rhombohedral. It is interesting to observe that, contrary to what is observed in tetragonal PZT, a to c domain switching is not visible in this experiment.

One can also compare this result with Landau-Ginzburg-Devonshire (LGD) theory prediction. Du et al. [8] deduced from LGD that the most favourable phase and orientation for single crystal / single domain PZT in order to maximize piezoelectricity is [100]-oriented rhombohedral in the MPB area. Here, our result suggests that for polycrystalline [100]-oriented 52/48 PZT, it is better to work with PZT in the tetragonal phase, but still very close to the MPB in order to have the propensity to transit easily to the rhombohedral phase when the electric field is applied. We suggest that our result could enable to improve even further $d_{33,eff}$ by adjusting PZT composition in order to exhibit 100% P4mm phase at 0V without losing the capability to induce its complete transition towards 100% rhombohedral phase. The whole transition from 100% P4mm to 100% R3m should induce $d_{33,eff}$ as high as 250 pm/V according to the peak positions method detailed in the previous section. That represents a 67%-increase compared with our experimental $d_{33,eff}$. Consequently, one should gently increase the amount of Ti into PZT in order for it to be slightly more tetragonal.

Finally, Fig. 3 suggests that one could obtain higher strain by using pure tetragonal PZT and transiting from 100% a-domains to 100% c-domains by applying voltage. However, Kobayashi et al. have recently observed that this complete a to c-domains transition in pure tetragonal $PZT_{30/70}$ requires large electric field (0.4 MV/cm) and induces cracks into PZT [20]. This result suggests that only PZT compositions in an MPB zone provide sufficient polarization mobility to take advantage of phase transitions to enhance piezoelectric properties.

In this paper, we observed by *in operando* X-Ray diffraction that a tetragonal to rhombohedral phase transition is experienced by MPB 52/48 PZT thin films when an external electric field is applied. Moreover, the field-induced domain switching in the tetragonal phase appeared to be negligible in these films. Besides, the films' piezoelectric coefficient $d_{33,eff}$ measured on the one hand with a piezoelectric set-up and on the other hand deduced from X-Ray peaks displacements versus the applied field appeared very similar. This proves that there is a strong correlation between the piezoelectric properties and the field-induced phase transition in MPB PZT films. This result suggests that one could probably improve even further $d_{33,eff}$ by favouring the field-induced phase transition notably by fine tuning PZT films composition.

**Acknowledgements**
The research described in this paper was supported by the French Ministry of Defense, DGA (Direction Générale de l'Armement).The authors acknowledge access to the Nanocharacterization Platform (PFNC) at Minatec Campus in Grenoble (France).